\documentclass[epj]{svjour}

\usepackage{graphicx}
\usepackage{fancyhdr}
\def\makeheadbox{{
 }}
\def\mytitle{My title} 
\def\myauthors{My name}  
\def\mytype{My type of session}
\def\mysession{My session}

\def\mytitle{SUSY Searches in All-Hadronic States with Large MET at the LHC} 
\def\myauthors{M. Tytgat}    
\def\mytype{Contributed Talk}    
\def\mysession{Colliders - SUSY Phenomenology}

\begin{document}
\title{SUSY Searches in All-Hadronic States with Large MET at the LHC}
\author{M. Tytgat 
\thanks{Also at the University of Gent, Belgium; \emph{Email:}
  michael.tytgat@cern.ch}
(on behalf of the ATLAS \& CMS Collaborations)
%
}                     
%
%
\institute{European Laboratory for Nuclear Research CERN, CH-1211 Geneva 23,
  Switzerland 
}
%
\date{}
\abstract{
The CMS \& ATLAS search strategy for SUSY in inclusive
multijet plus high missing transverse
energy final states is reviewed. This canonical SUSY signature may be a viable
discovery channel for low mass SUSY in the early phase of the LHC.
Methods for Standard Model background estimates, MET studies and filters for 
instrumental background are presented.
} 
\maketitle
\section{Introduction}
\label{sec:intro}
Supersymmetry (SUSY) is one of the most robust extensions of
the Standard Model. 
Several compelling arguments are in favor of supersymmetry at the TeV
scale, which should enable its discovery at the LHC. The simplest
supersymmetrization of the Stan\-dard \linebreak
Mod\-el is known as the Min\-i\-mal
Super\-sym\-met\-ric Stan\-dard Mod\-el (MSSM)~\cite{ref:SUSY}. 
However, because of the large number of
free parameter of this model, SUSY 
analysis studies are usually carried out in more
constrained models which make certain assumptions on the SUSY
breaking mechanism.
Inclusive searches are mainly \linebreak
done in the framework of the min\-i\-mal
Su\-per\-grav\-i\-ty (mSUGRA) model, which has only five independent 
parameters~: the common gaugino mass the GUT scale $m_{1/2}$, the common
scalar mass at the GUT scale $m_0$, the common trilinear coupling at the GUT
scale $A_0$, the ratio of the vacuum expectation values of the two Higgs
doublets $\tan\beta$ and the sign of the Higgsino mixing parameter 
$sign(\mu)$. If $R$-parity is conserved, the stable neutral
lightest supersymmetric 
particle (LSP),
which remains undetectable, will lead to SUSY event signatures with large
missing energy. 
For their SUSY studies, both ATLAS and CMS have chozen a set of benchmark 
points to cover as much as possible the different event signatures that may 
occur in the different regions of the mSUGRA parameter space. 

The cross section for sparticle production at the LHC is
usually dominated by squark and gluino production. Their cascade decay results
in final states characterized by multiple hard jets plus large $E_T^{miss}$
plus often also one or more leptons~\cite{ref:baer}.
Here, the main parts of the ATLAS and CMS
inclusive SUSY searches in all-hadronic states, i.e. no leptons, will be 
discussed. More information on these analysis studies can be \linebreak
found in
\cite{ref:atlasptdr,ref:atlascsc,ref:cmsptdr2}. 
 
\section{Missing Transverse Energy}
\label{sec:met} 
Missing $E_T$ is a powerful tool for SUSY discovery, but it is also a complex
object. Apart from real missing $E_T$ due to undetected particles, it will 
include contributions from non-collision background such as beam halo or
cosmic muons and from detector effects like instrumental noise, hot or dead
channels or cracks in between different parts of the detector. For SUSY
searches it is important to understand and remove such fake $E_T^{miss}$ 
contributions, especially in the high tails of the distribution where SUSY
signals are expected.

\begin{figure}[ht]
\begin{center}
\includegraphics[width=.5\textwidth]{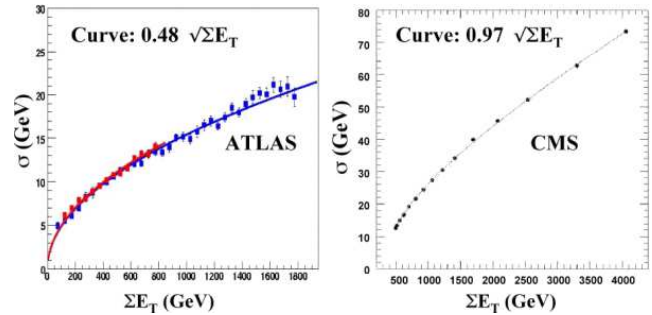}
\end{center}
\caption{The ATLAS (left) and CMS (right)
Missing Transverse Energy resolution as function 
of the total
  transverse energy ($\Sigma E_T$) in the event for QCD jet production.}
\label{fig:met} 
\end{figure}
The missing $E_T$ resolution is observed to depend on the overall activity of
the event, which is characterized by the scalar sum of the transverse energy
in all calorimeter cells, $\Sigma E_T$. 
Figure~\ref{fig:met} displays the expected 
$E_T^{miss}$ resolution as function of 
$\Sigma E_T$, obtained for 
QCD jet production for both the ATLAS and CMS detectors. 
The $E_T^{miss}$
resolution for both experiments is expected to be dominated by the respective 
calorimeter resolutions.

\begin{figure}[ht]
\begin{center}
\includegraphics[width=.24\textwidth]{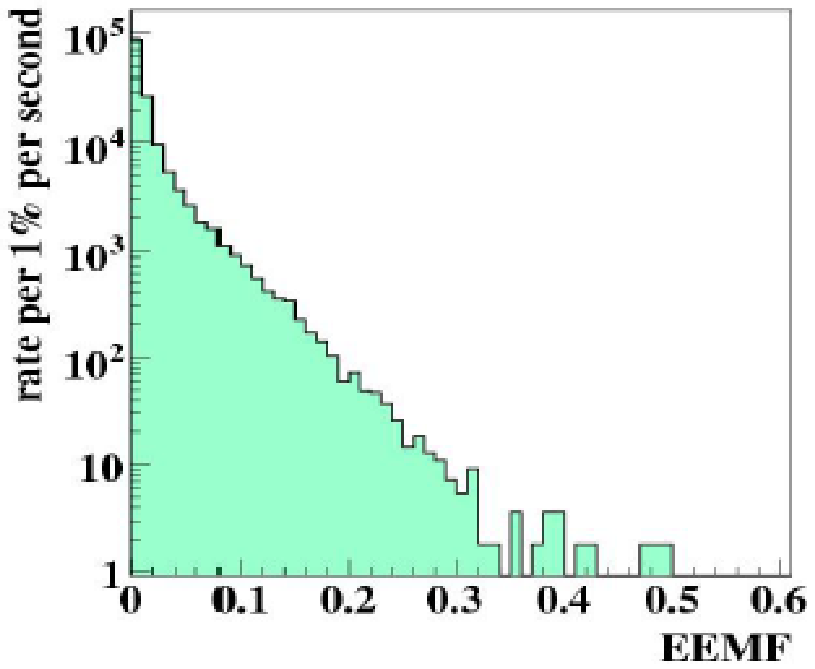}\hfill
\includegraphics[width=.24\textwidth]{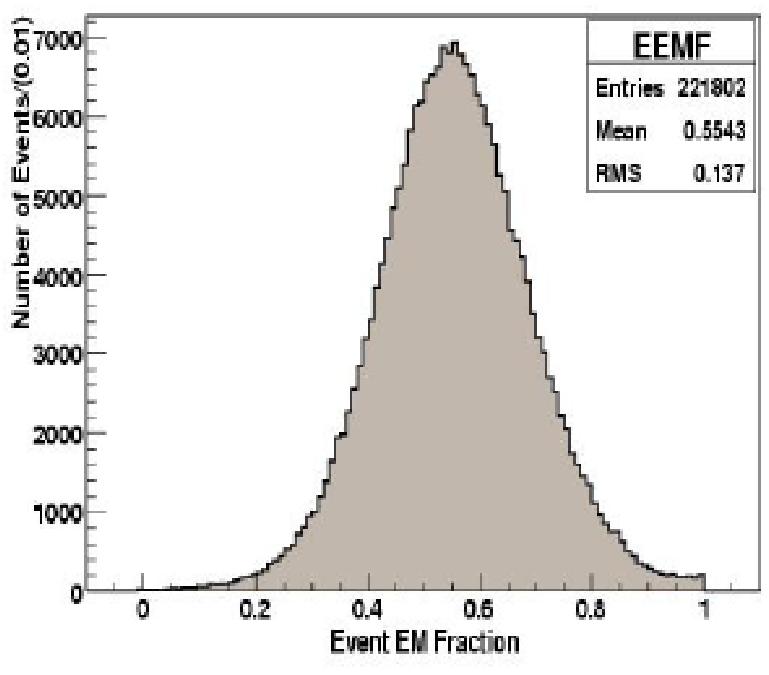}
\end{center}
\caption{The $F_{em}$ for beam halo events in CMS (left) and for a $t\bar{t}$
  event sample (right).}
\label{fig:eemf}
\end{figure}
An $E_T^{miss}$ cleanup procedure used by CMS is based on the event 
electromagnetic fraction
\begin{equation}
F_{em}=\frac{\sum_{j=1}^{N_{jet}}P_{Tj}\times
  f_{em}^j}{\sum_{j=1}^{N_{jet}}P_{Tj}},
\end{equation}
for jets within the calorimeter acceptance $|\eta|\le 3$ and 
with $f_{em}^j$ the jet electromagnetic fraction, 
and on the event charged fraction
\begin{equation}
F_{ch}=\left.\left<\frac{(\sum_i^{tracks}P_{Ti})_j}{P_{Tj}}\right>\right|_{N_{jet}},\nonumber
\end{equation}
where the average is taken over jets within $|\eta|<1.7$ and the sum for every
jet runs over 
the particle tracks that can be matched in 
$(\eta,\phi)$ space to that jet. This procedure is
promising to remove cosmic ray events and
accelerator- and detector-related backgrounds, like beam halo or electronic
noise. The $f_{em}$ of jets in 
cosmic or beam halo events will, depending on where the particles hit the
electromagnetic or hadronic calorimeters,
differ from the typical $f_{em}$ for
jets in normal collisional events. Also, the jet charged
fractions in such events will be different due to the fact that very 
little or no particle tracks can be associated to them.
Figure~\ref{fig:eemf} shows the $F_{em}$ for simulated beam
halo events in the CMS detector and for a $t\bar{t}$ event sample for
comparison. Requiring at least one primary vertex in the event and e.g.
$F_{em}>0.1$ 
and $F_{ch}>0.175$ removes most of the 
background, while retaining $\sim 91$~\% of the signal
in a SUSY event sample at the CMS LM1 benchmark point.   

\section{Standard Model Backgrounds}
\label{sec:SM}
Finding a SUSY signal requires not only a thorough understanding of the 
detector,
but also a deep knowledge of all possible
Standard Model background sour\-ces. To minimize systematic uncertainties from 
Monte Carlo model predictions, both ATLAS and CMS are studying methods to 
estimate
backgrounds in SUSY searches directly from the data.
Most important Standard Model 
background sources are QCD multi-jet production, top quark
pairs, $Z/W +$ jets and diboson plus jets production.

In the case of e.g.~$W$ 
and $t\bar{t}$ production, high $E_T^{miss}$ is normally
associated to the presence of leptons in the final state, such that the
all-hadronic channel will suffer from this type of 
background events only if the
lepton identification fails. In addition, muons in contrast to electrons, 
leave very little energy in the
calorimeters, so that in general 
they will also contribute to the $E_T^{miss}$ if the muon
identification fails.
   
\subsection{QCD Jet Production}
\label{sec:qcd}
The very high cross section makes QCD jet production the dominant Standard
Model background source in a multi-jet plus large missing transverse energy
data-sample. The missing transverse energy in QCD events is largely due to jet
mis-measurements and detector resolution. The large cross section in
combination with the trigger and data-acquisition bandwidth restrictions make
it difficult to collect QCD datasets with low $E_T$ thresholds and 
the extraction of shapes and normalizations for QCD
background in SUSY searches will only be possible with prescaled triggers.
However, for large missing transverse energy,
jet mis-measurements are likely to pull the $E_T^{miss}$ direction close in
$\phi$ to the mis-measured jet direction. This means that topological
requirements can be used to suppress QCD background contributions.

\begin{figure}[h]
\begin{center}
\includegraphics[width=.24\textwidth]{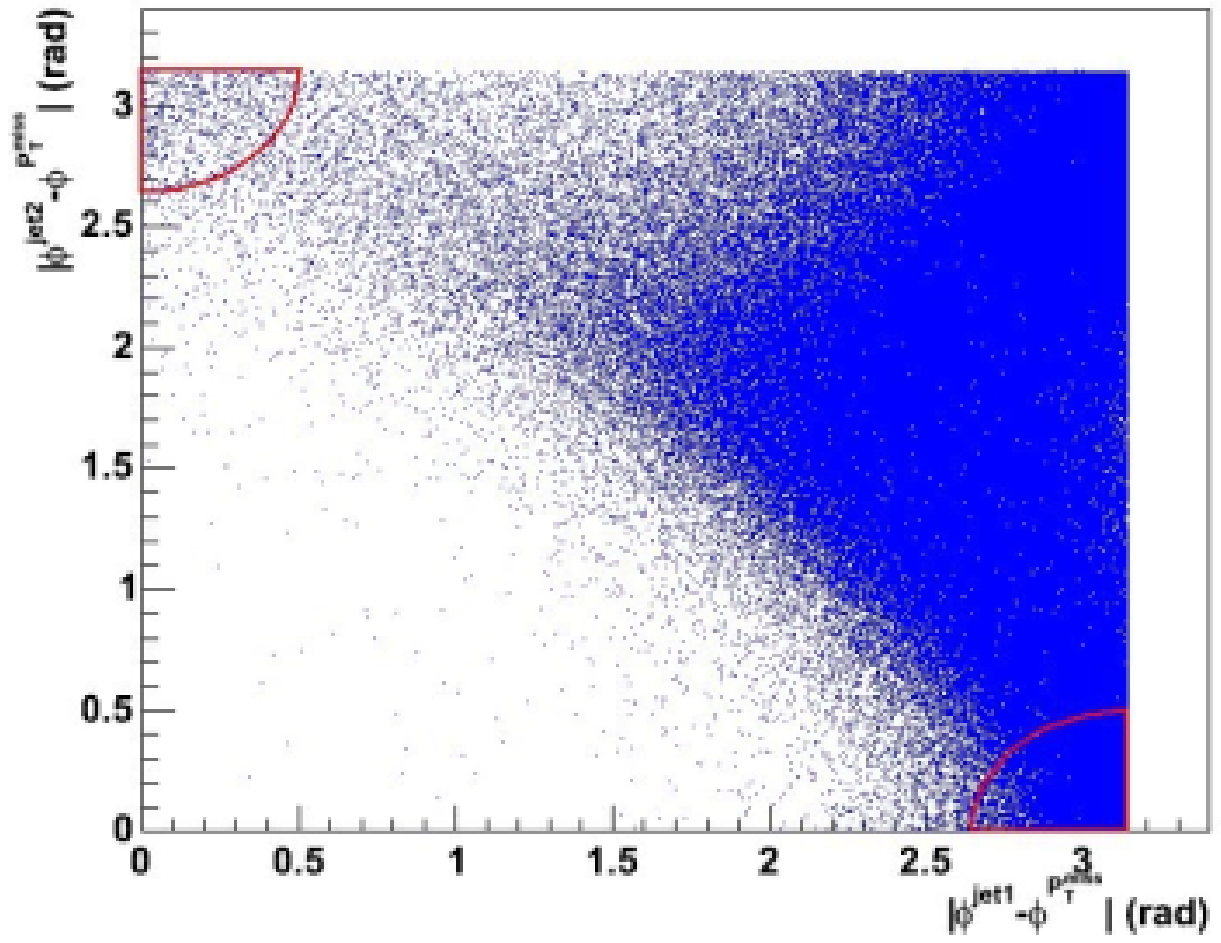}\hfill
\includegraphics[width=.24\textwidth]{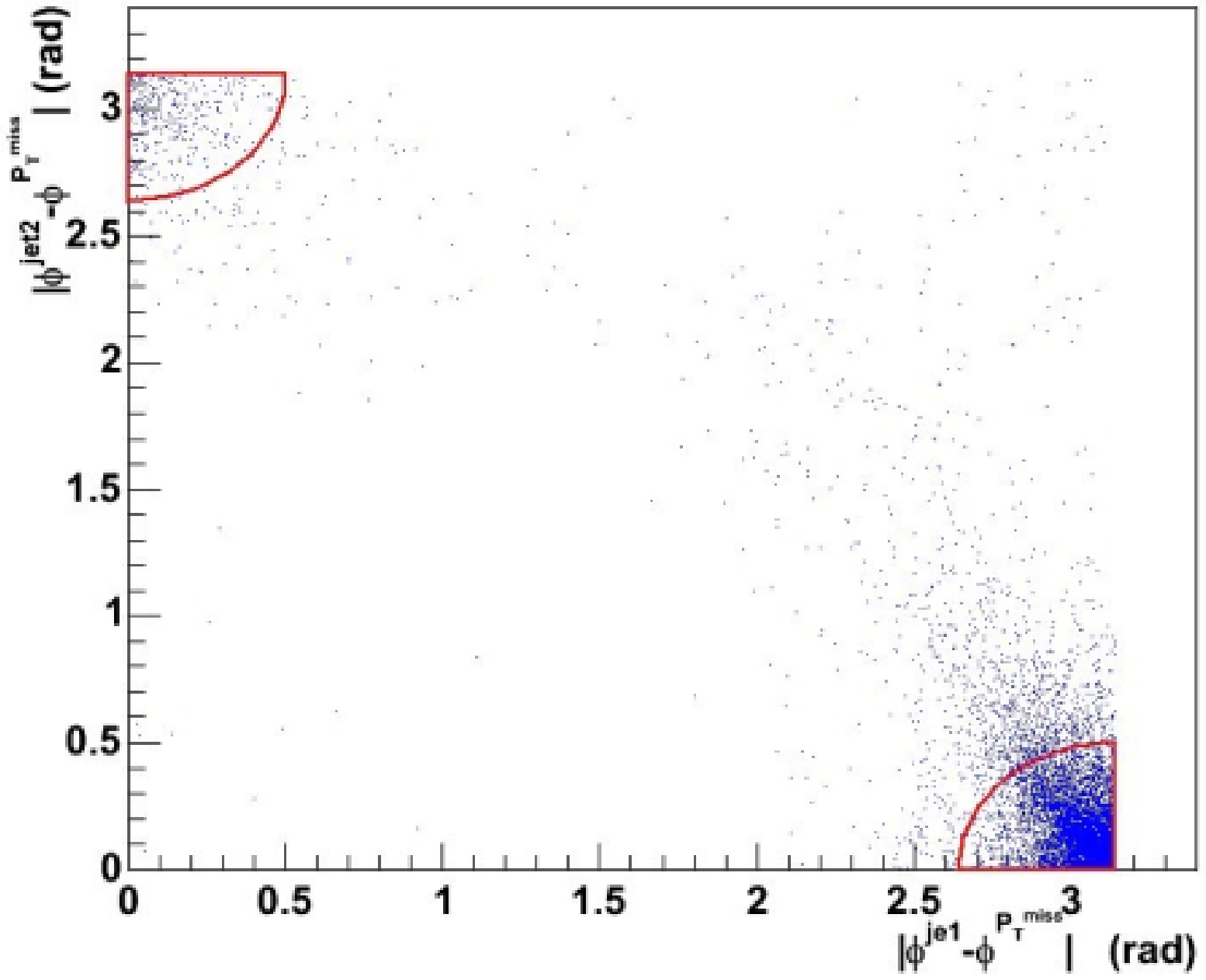}
\end{center}
\caption{$\delta\phi_1$ versus $\delta\phi_2$ for SUSY signal (left) and QCD
  di-jet events (right) for events with $R_{1,2}>0.5$.}
\label{fig:QCD1}
\end{figure}

\begin{figure}[h]
\begin{center}
\includegraphics[width=.24\textwidth]{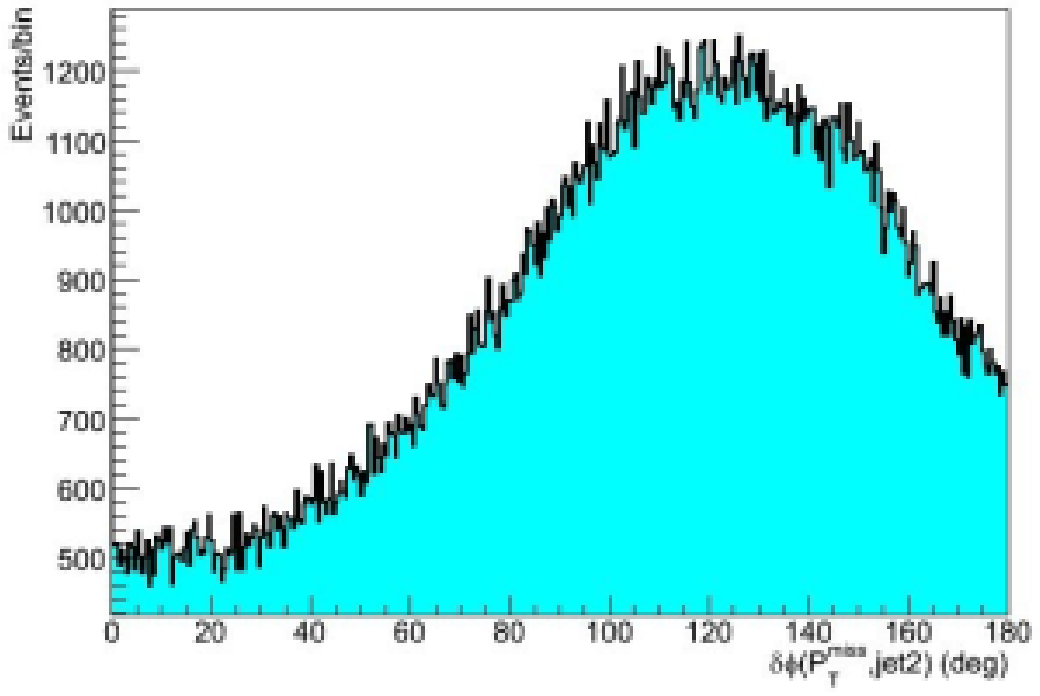}\hfill
\includegraphics[width=.24\textwidth]{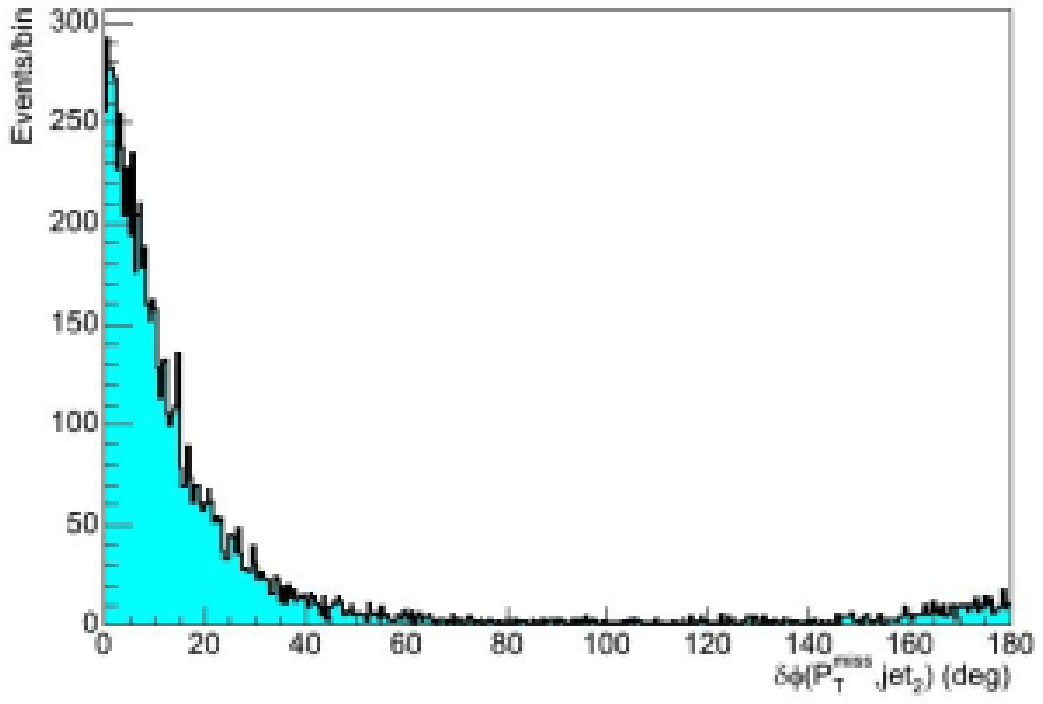}
\end{center}
\caption{$\delta\phi_2$ for SUSY signal (left) and QCD di-jet events (right).}
\label{fig:QCD2}
\end{figure}

CMS uses the correlation between 
$\delta\phi_1=|\phi_{j(1))}-\phi(E_T^{miss})|$ and 
$\delta\phi_2=|\phi_{j(2))}-\phi(E_T^{miss})|$ with $\phi_{j(1,2)}$ being the
$\phi$ angle of the first and second leading jet in the event.
In a first step, only events with
$R_{1(2)}=\sqrt{\delta\phi_{2(1)}^2+(\pi-\delta\phi_{1(2)})^2}>0.5$ are 
accepted.  
Figure~\ref{fig:QCD1} shows $\delta\phi_1$ versus $\delta\phi_2$ 
for QCD di-jet and SUSY signal events at the CMS LM1 benchmark point.
In addition no jet in the event should be closer than 0.3~rad to the
$E_T^{miss}$ direction and the second leading jet should be further than
$20^{\circ}$ from it as illustrated in 
Figure~\ref{fig:QCD2}. Together with an event selection of $N_{jet}\ge 2$ and
$E_T^{miss}>93$~GeV these angular requirements retain $\sim 90$~\% of
the SUSY signal at the CMS LM1 benchmark point, while rejecting $\sim 85$~\% 
of the QCD events. 

\subsection{$W/Z$ Boson Production}
\label{sec:wz}
$W(\rightarrow l\nu)$ and $Z(\rightarrow\nu\bar{\nu})$ boson production in 
association with jets will also give rise to
final states with large $E_T^{miss}$ plus multiple jets. Both ATLAS and CMS
are developing methods to estimate all such contributions by combining Monte
Carlo prediction with measured data to reduce systematic uncertainties due to
e.g. the QCD renormalization scale, 
the choice of parton density function or jet energy scale.

A technique used by CMS relies on the fact that the $Z+N$ jets cross section
is proportional to $\alpha^N_s$, such that the ratio of the number of events
in adjacent jet multiplicity bins is expected to be constant and proportional
to $\alpha_s$. In this way, the measured $Z(\rightarrow\mu\mu)+2$ jets rate 
can be
used to normalize the Monte Carlo predictions for $Z+\ge 3$ jets via the 
measured $R\equiv\frac{\mbox{d}N_{events}}{\mbox{d}N_{jets}}$ ratio. In 
addition, 
the ratio $\rho\equiv\frac{\sigma(pp\rightarrow
  W(\rightarrow\mu\nu)+jets)}{\sigma(pp\rightarrow
  Z(\rightarrow\mu\mu)+jets)}$ can then be used to normalize the $W+jets$
Monte Carlo predictions. For 1~fb$^{-1}$ a systematic uncertainty of 
$\sim 5$~\% is obtained, which is dominated by the luminosity measurement and 
the uncertainties on the measured $R$ and $\rho$.

\subsection{Top Quark Pair Production}
\label{sec:top}
Due to its relatively large cross section and its typical multi-jet, high
$E_T^{miss}$ and leptonic final state, top quark pair production is a
particular important background source for SUSY searches and therefore, 
estimating the top background directly from measured data is essential. 
Especially for the early
data, both ATLAS and CMS are looking into reconstructing the $t\bar{t}$ 
channel without $b$-tagging. 

\begin{figure}[h]
\begin{center}
\includegraphics[width=.24\textwidth]{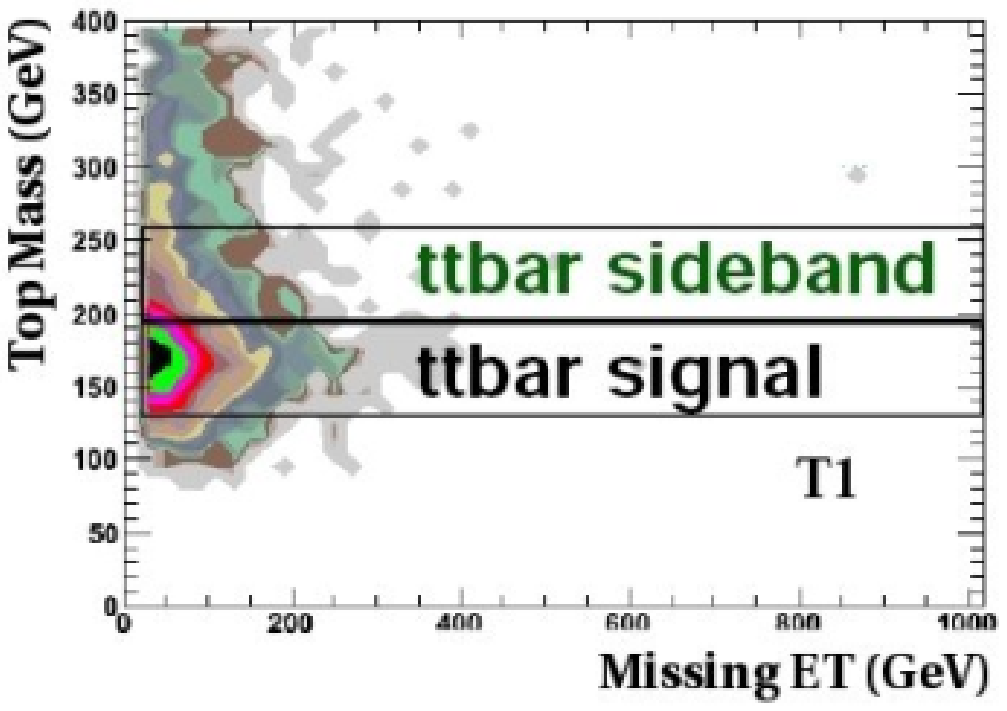}\hfill
\includegraphics[width=.24\textwidth]{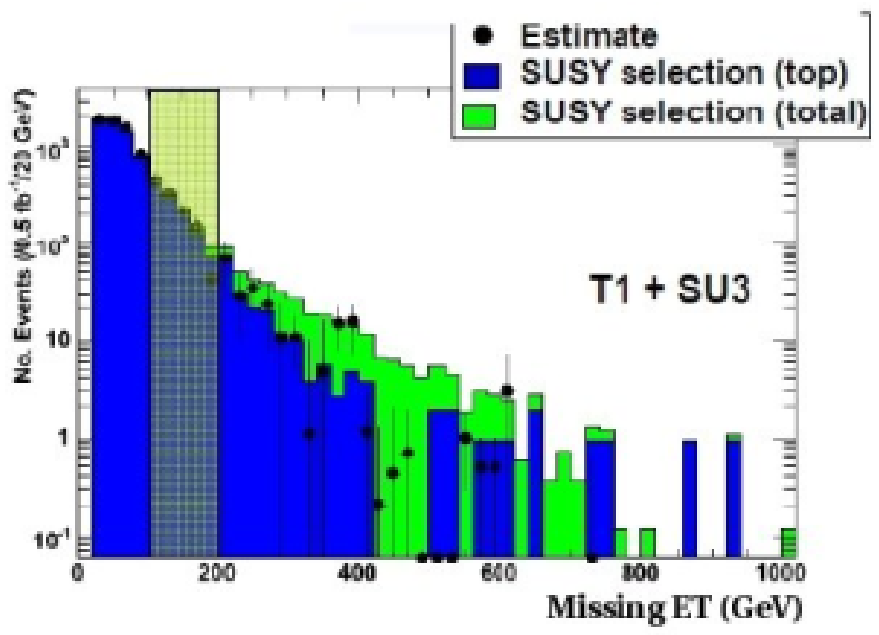}
\end{center}
\caption{Reconstructed top invariant mass versus $E_T^{miss}$ for a $t\bar{t}$
sample (left) and $E_T^{miss}$ for a $t\bar{t}$ and SUSY plus $t\bar{t}$
sample together with the normalized, 
estimated $t\bar{t}$ distribution (right).}
\label{fig:top}
\end{figure}

A procedure used by ATLAS to estimate the $t\bar{t}$ contamination in SUSY
analyses is illustrated in the left panel of 
Figure~\ref{fig:top}, where the
reconstructed top invariant mass (no $b$-tagging used) 
for the semi-leptonic channel is displayed 
versus $E_T^{miss}$. An estimation of the $E_T^{miss}$ distribution for 
$t\bar{t}$ production
is determined in a signal region around the top quark mass and corrected for
combinatorial background by subtracting the corresponding $E_T^{miss}$
distribution taken from a top quark sideband region. The right panel of
Figure~\ref{fig:top} shows the $E_T^{miss}$ distribution for a $t\bar{t}$ plus
SUSY Monte Carlo event sample. By normalizing the estimated $t\bar{t}$
distribution to the total distribution in the low $E_T^{miss}$
region, where the SUSY signal is small, one effectively obtains a fair 
description of the overall $t\bar{t}$ distribution also at high $E_T^{miss}$.

\section{All-Hadronic SUSY Event Selection and Results} 
\label{sec:cuts}
For the inclusive SUSY searches in
all-hadronic states the typical 
event selection criteria used by
ATLAS are $N_{jet}\ge 4$, $P_T^{j1}>100$~GeV, $P_T^{j4}>50$~GeV,
$E_T^{miss}>100$~GeV, zero reconstructed leptons and transverse sphericity
$S_T>0.2$ to suppress e.g. QCD di-jet events. 

The corresponding event 
selection in
the CMS analysis is $N_{jet}\ge 3$, $E_T^{j1}>180$~GeV, $E_T^{j2}>110$~GeV,
$E_T^{j3}>30$~GeV, $E_T^{miss}>200$~GeV plus additional 
$E_T^{miss}$ clean-up criteria (see Section~\ref{sec:met}),
topological cuts on $\Delta\phi$ between 
the jets and the $E_T^{miss}$ direction for QCD jet suppression 
(see Section~\ref{sec:qcd}) and
$H_T=E_T^{j2}+E_T^{j3}+E_T^{j4}+E_T^{miss}>500$~GeV. Instead of using explicit
lepton identification, CMS uses an indirect lepton veto to reject $W$, $Z$ and
$t\bar{t}$ backgrounds, while retaining as much as possible the SUSY signal. 
This lepton veto removes events with high-energy electrons or muons and 
consists of two parts~: events are
only accepted if the first and second leading jet are not purely
electromagnetic ($f_{em}<0.9$) and if the leading high-$P_T$ track in the 
event satisfies
a certain non-isolated criterium. 

\begin{figure*}
\begin{center}
\includegraphics[width=.4\textwidth]{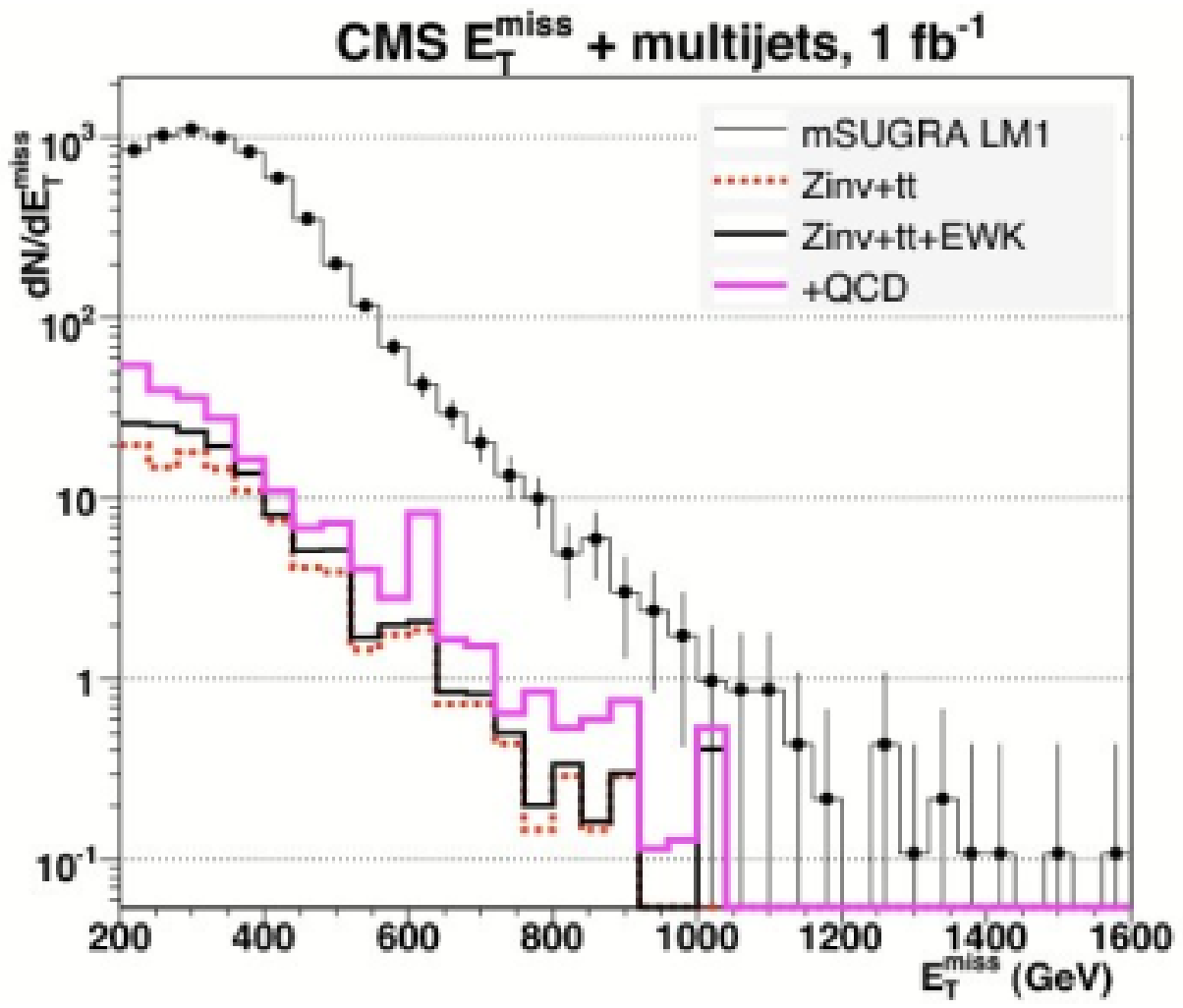}\hfill
\includegraphics[width=.49\textwidth]{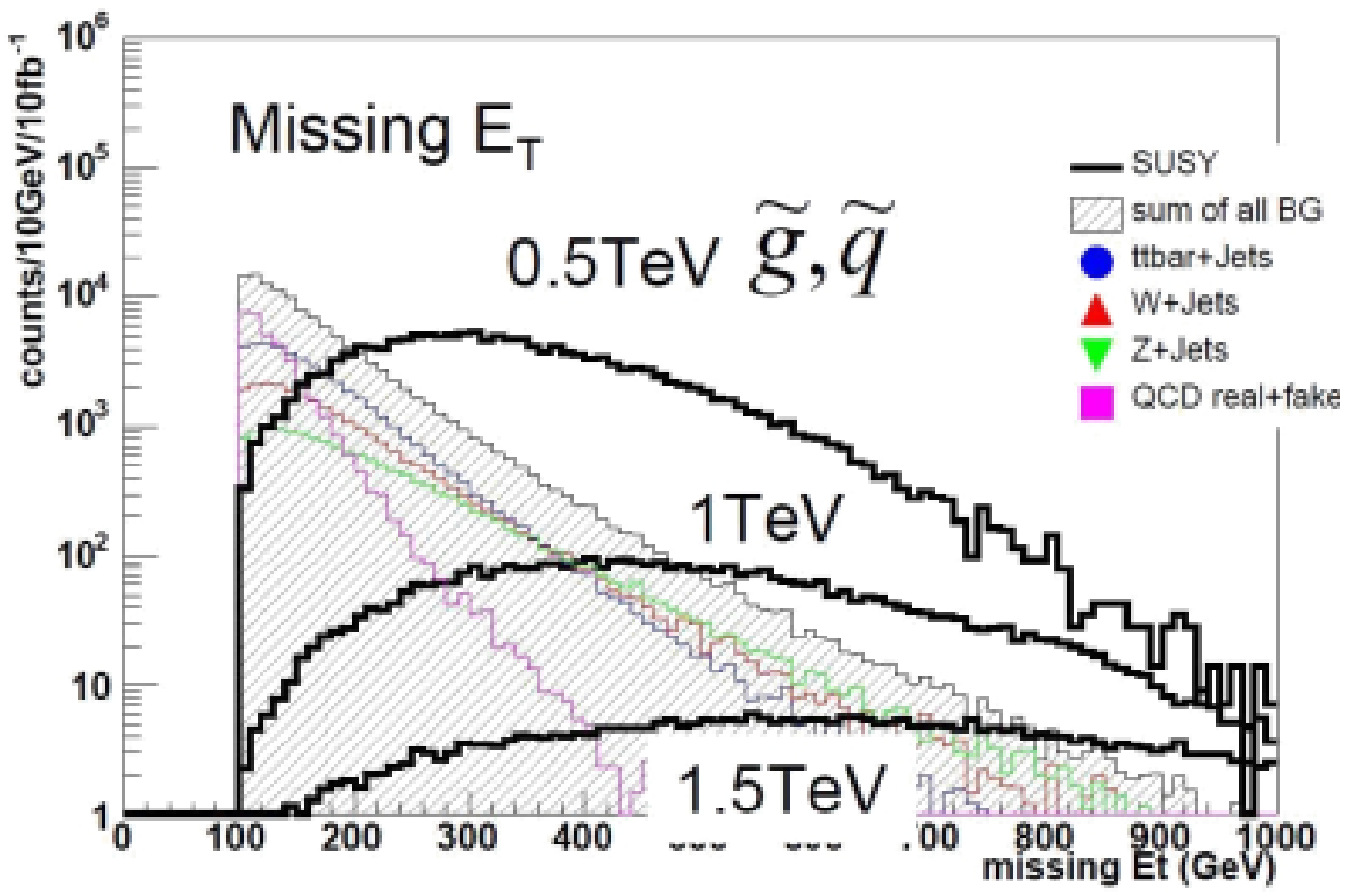}
\end{center}
\caption{$E_T^{miss}$ distributions in all-hadronic SUSY searches for CMS
  (left) and
  ATLAS (right) with 1~fb$^{-1}$.}
\label{fig:res}
\end{figure*}
Figure~\ref{fig:res} displays the resulting 
$E_T^{miss}$ distribution for multi-jet plus
high $E_T^{miss}$ SUSY searches with 1~fb$^{-1}$ for both CMS and ATLAS after 
the
abovementioned event selections. The CMS result is given at the LM1 benchmark
point, with $m(\tilde{g})\approx 600$~GeV and $m(\tilde{q})\approx 550$~GeV, 
while the ATLAS result is given for $\tilde{g}$ and $\tilde{q}$ mass scales of
0.5, 1.0, 1.5~TeV. For the low mass scale, both experiments expect a clear
excess of events above the estimated Standard Model background.

\section{mSUGRA Discovery Reach}
\label{sec:reach}

\begin{figure*}
\begin{center}
\includegraphics[width=.45\textwidth]{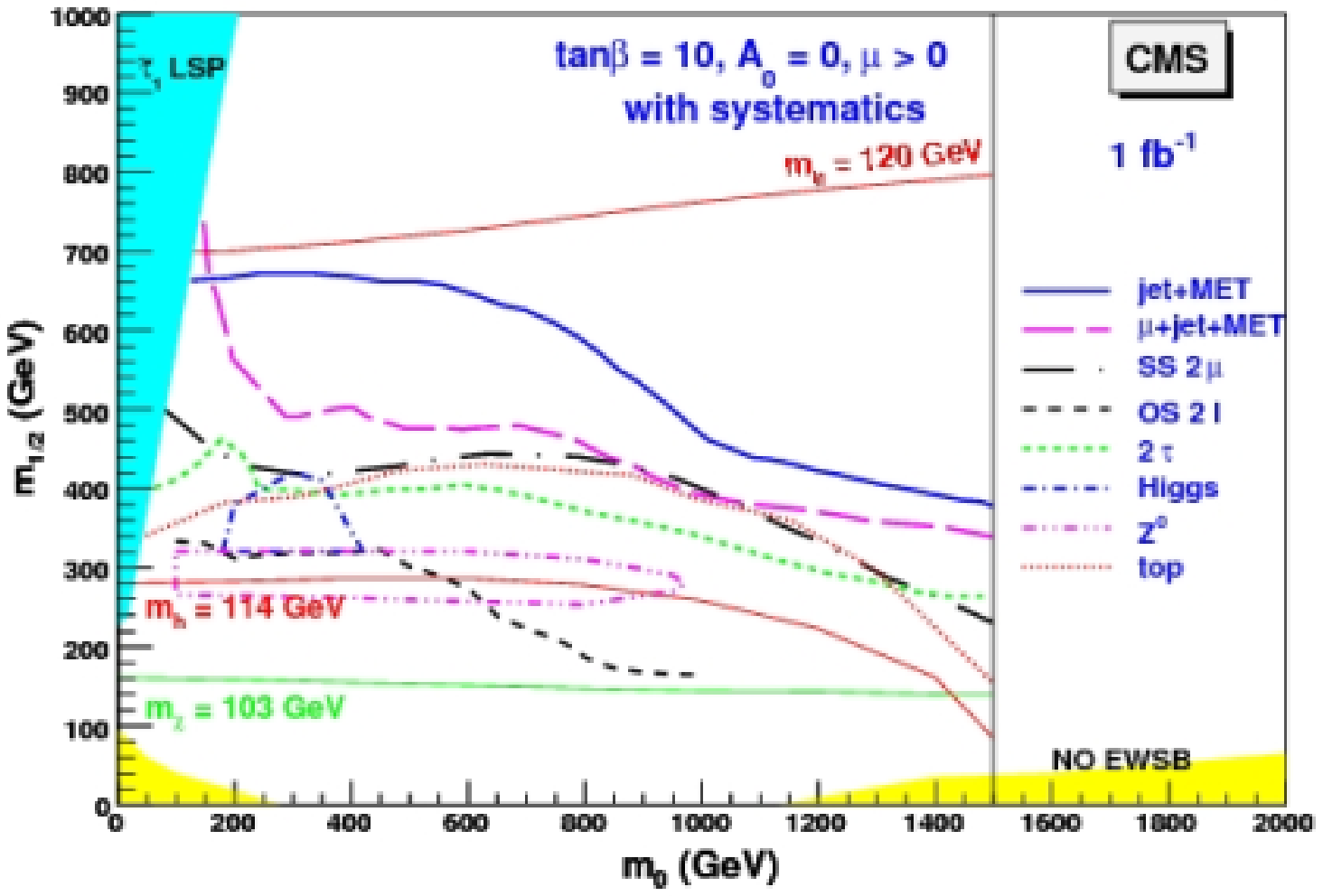}\hfill
\includegraphics[width=.5\textwidth]{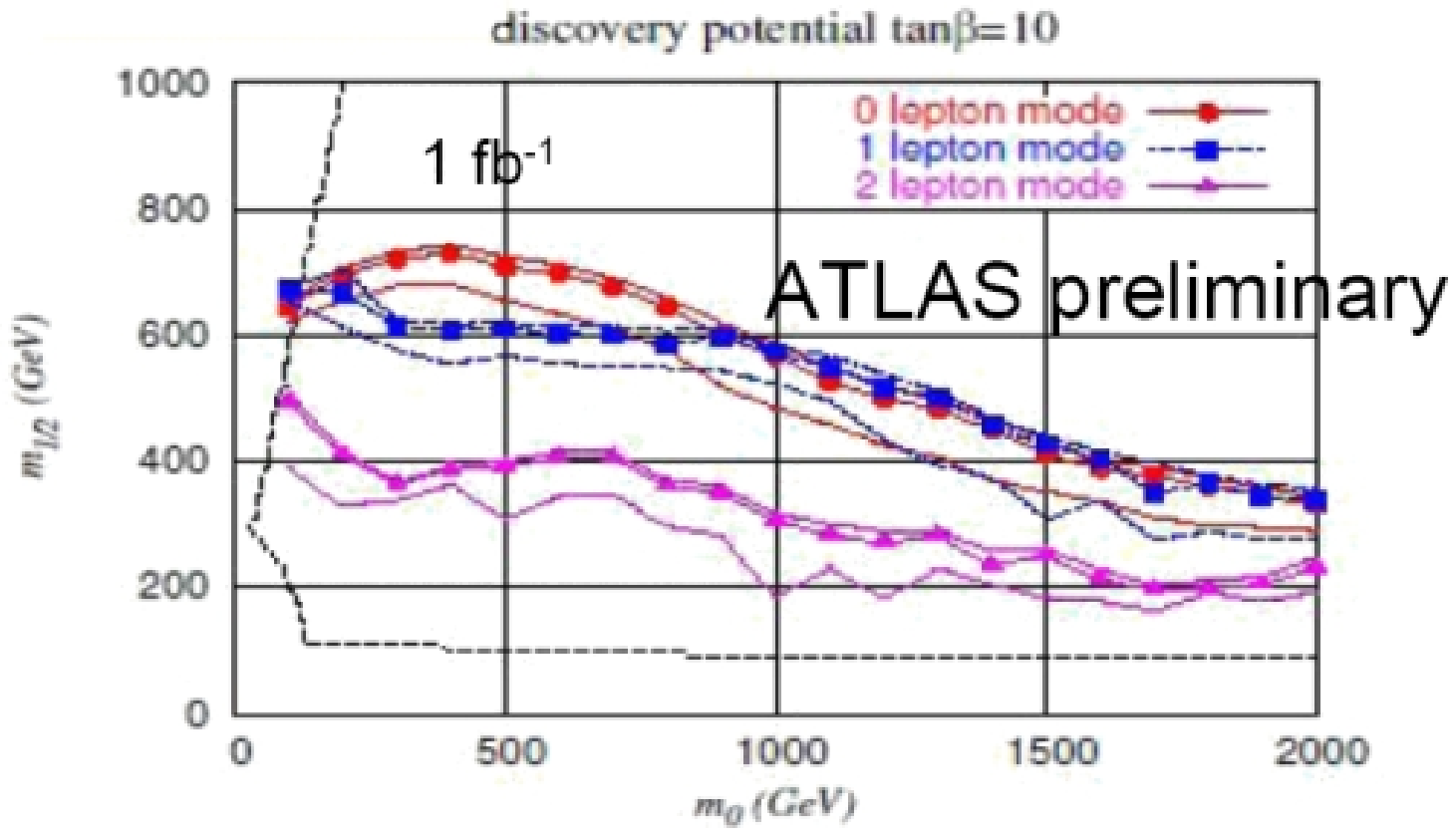}
\end{center}
\caption{$5\sigma$ mSUGRA discovery reaches for different channels for
  1~fb$^{-1}$ for CMS (left) and ATLAS (right).}
\label{fig:reaches}
\end{figure*}

Both ATLAS and CMS have performed a scan of the mSUGRA $(m_0,m_{1/2})$ plane.
Figure~\ref{fig:reaches} shows the $5\sigma$ reach contours for 1~fb$^{-1}$
for both experiments, which for the all-hadronic searches appear quite 
comparable. The scans
demonstrate that with 1~fb$^{-1}$ all of the low mass region for
$\tan\beta=10$, $A_0=0$ and $\mu>0$ can be observed.
The reach contours 
are also shown for other SUSY analysis studies. While these are also very 
promising,
it is clear that the inclusive all-hadronic searches yield the best SUSY 
discovery
potential for the early LHC data. 

\section{Conclusions}
\label{sec:concl}
Using the inclusive multi-jet plus high $E_T^{miss}$ event signature an 
early discovery of SUSY at the LHC should be possible if the latter would
manifest itself at a low mass scale. 
Both the ATLAS and CMS analysis predict a comparable discovery reach in the
mSUGRA parameter space for this all-hadronic channel. Of particular importance
for SUSY discovery is a deep understanding of the observed missing transverse 
energy and correct estimates of the different Standard Model background
sources, where data-driven methods are actively being studied by both
collaborations.  
    
\section{Acknowledgements}
The author would like to thank S.~Asai and D.~Tovey for providing the ATLAS
results included in this contribution. 

%

\begin{thebibliography}{999}
%
%
\bibitem{ref:SUSY} H.P.~Nilles, Phys. Rep. \textbf{110}, (1984) 1 and
  references therein.
\bibitem{ref:baer} H.~Baer, C.H.~Chen, F.~Paige and X.~Tata, Phys. Rev. D
  \textbf{52}, (1995) 2746; Phys. Rev. D \textbf{53}, (1996) 6241.
\bibitem{ref:atlasptdr} The ATLAS Collaboration, CERN-LHCC-99-15 (1999). 
\bibitem{ref:atlascsc} The ATLAS Collaboration, ATLAS CSC Notes, 
{\it in preparation.}
\bibitem{ref:cmsptdr2} The CMS Collaboration, CERN-LHCC-2006-021 (2006);
  J. Phys. G: Nucl. Part. Phys. \textbf{34}, (2007) 995-1574.

\end{thebibliography}
%

\end{document}